\documentclass[runningheads]{llncs}
\usepackage[T1]{fontenc}
\usepackage{graphicx}
\begin{document}
\title{DAnTE: a taxonomy for the automation degree of software engineering tasks}

\author{Jorge Melegati\inst{1}\orcidID{0000-0003-1303-4173} \and
Eduardo Guerra\inst{1}\orcidID{0000-0001-5555-3487}
}

\institute{Free University of Bozen-Bolzano, Bolzano, Italy %
\email{\{jorge.melegati,eduardo.guerra\}@unibz.it}\\
}
\maketitle              
\begin{abstract}
Software engineering researchers and practitioners have pursued manners to reduce the amount of time and effort required to develop code and increase productivity since the emergence of the discipline. Generative language models are just another step in this journey, but it will probably not be the last one. In this chapter, we propose DAnTE, a Degree of Automation Taxonomy for software Engineering, describing several levels of automation based on the idiosyncrasies of the field. Based on the taxonomy, we evaluated several tools used in the past and in the present for software engineering practices. Then, we give particular attention to AI-based tools, including generative language models, discussing how they are located within the proposed taxonomy, and reasoning about possible limitations they currently have. Based on this analysis, we discuss what novel tools could emerge in the middle and long term.

\keywords{software engineering automation  \and AI for software engineering \and generative AI.}
\end{abstract}

\section{Introduction}
\label{sec:introduction}

Human history is defined by the use of tools. From the use of rudimentary stones for preparing food in the Paleolithic to rockets taking us to space today, tools have been used to augment our capabilities or even to give us newer and more complex ones. As with any other human activity, software engineering (SE) has also experienced, since its inception, the continuous creation of tools to increase productivity, reduce errors, and facilitate the work of those involved in the activity. High-level programming languages, integrated development environments (IDEs), and frameworks, just to name a few, allowed the development of increasingly complex software systems with reduced effort and time spent. The emergence of a novel generation of tools based on artificial intelligence (AI), driven by generative large language models (LLMs), promises to be a disruption to how software is developed, intensifying this tendency. However, a framework to understand the evolution of these tools, including generative AI, still lacks. In this chapter, based on a literature review of taxonomies in SE and other fields, we propose DAnTE, a taxonomy of SE automation tools that considers the whole evolution of these tools, positioning generative AI-based tools in a historical continuum. 

DAnTE consists of six levels of degree of automation: No automation (Level 0), Informer (Level 1), Suggester (Level 2), Local generator (Level 3), Global generator (Level 4), and Full generator (Level 5). To evaluate the taxonomy, we use it to classify two sets of tools focused on different SE activities: coding and testing. This utility demonstration indicates that the taxonomy could be used for classifying different tools for the automation of diverse SE tasks with clear advantages over the existent classification schemes described in the literature. Practitioners could employ it for evaluating and choosing tools to be used in different contexts. The taxonomy could be also useful for classifying research endeavors and guiding the development of novel tools. 

\section{Related Work}
\label{sec:related_work}

A taxonomy is a classification system defined by a collection of classes~\cite{Usman2017,Ralph2019}. A class is a set of properties shared by a set of instances~\cite{Parsons2008,Ralph2019}, where ``an instance can be a material object, action, event, or any other phenomenon''~\cite{Parsons2008}. Taxonomies are useful since they serve two functions: cognitive efficiency and inference support~\cite{Parsons2008,Ralph2019}. Cognitive efficiency regards the possibility of remembering features of classes (a smaller number) instead of instances (a larger number)~\cite{Parsons2008}. For example, a class ``cat'' already describes a set of characteristics, such as having whiskers and meowing, so we do not need to store the information that each cat we know has whiskers and meows, we only store that a specific cat is an instance of the class ``cat'' and, consequently, has whiskers and meows. Inference regards the possibility of deducing unobserved properties of an instance based on the class it pertains or ``about other (concurrent) phenomena, about possible future states of us and our environment''~\cite{Parsons2008}. Following on the previous example, when we got to know that a particular animal is a cat, we can already infer, based on the characteristics of the class, that it has whiskers and meows, for example.      

In the SE field, taxonomies are useful because ``they 
provide concepts and technical language needed for precise communication and education''. Specifically, a taxonomy for the automation degree of tasks could help practitioners to choose tools to be used in their projects, researchers to develop novel solutions, and educators to prepare the future actors for these tasks. 

In the SE literature, there are taxonomies focused on related topics. Feldt et al.~\cite{Feldt2018} proposed AI-SEAL, a taxonomy for the use of AI in SE. It consists of three facets: point of application (PA), type of AI applied (TAI), and level of automation (LA). PA regards ``when'' and ``on what'' the tool is being applied, and the possible values are process, product, or runtime. TAI represents the type of AI technology used, for instance, connectionist or symbolist. Finally, the level of automation, described in Table~\ref{tab:aiseal}, regards to which extent the tool assists the human in performing the task. The scale is adapted from the 10 levels of automation in Human-Computer Interaction proposed by Sheridan and Verplank~\cite{Sheridan2005}. The adoption of a taxonomy proposed for another field has the positive aspect of having been evaluated, however, even if it is related to the topic, it does not consider the specificalities of the new field, in our case, SE. Another issue, in our opinion, is the focus on AI, neglecting the long history of tools developed for SE automation.

\begin{table}
\renewcommand{\arraystretch}{1.2}
\centering
\caption{Levels of automation used in the AI-SEAL taxonomy~\cite{Feldt2018}}\label{tab:aiseal}
\begin{tabular}{c|p{0.9\textwidth}}
\hline
Level & Description \\
\hline
1 & Human considers alternatives, makes and implements decision. \\
2 & Computer offers a set of alternatives which human may ignore in making decision. \\
3 & Computer offers a restricted set of alternatives, and human decides which to implement. \\
4 & Computer offers a restricted set of alternatives and suggests one, but human still makes and implements final decision. \\
5 & Computer offers a restricted set of alternatives and suggests one, which it will implement if human approve. \\
6 & Computer makes decision but gives human option to veto before implementation. \\
7 & Computer makes and implements decision, but must inform human after the fact. \\
8 & Computer makes and implements decision, and informs human only if asked to. \\
9 & Computer makes and implements decision, and informs human only if it feels this is warranted. \\
10 & Computer makes and implements decision if it feels it should, and informs human only if it feels this is warranted.\\
\hline
\end{tabular}
\end{table}

Savary-Leblanc et al.~\cite{SavaryLeblanc2023} performed a systematic mapping study on the employment of software assistants in SE. They classified assistants into three types: informer systems, passive recommender systems, and active recommender systems. Informer systems simply display the results of data analysis wihtout any side effect. Passive recommender systems analyze data and potentially produce one or several alternatives for a decision-making problem. Finally, an active recommender system could also implement the decision. As examples of active recommender systems, the paper refers to tools that suggest code completion and code refactoring but also code documentation. Given the recent advances of generative LLMs, this class might accommodate several tools that have really different capabilities, such as code completion tools based on stemming and full-fledged code generation solutions based on LLMs.

Another related model is the Stairway to Heaven proposed by Olsson et al.~\cite{Olsson2012} to describe the transition to continuous deployment of software. The model consists of five stages: traditional development, agile R\&D organization, continuous integration, continuous deployment, and R\&D as an experiment system. At the final stage, the team ``acts based on instant customer feedback'' and ``software functionality is seen as a way of experimenting what the customer needs.'' In a follow-on paper~\cite{Bosch2018}, the same authors proposed three steps: requirement-driven, outcome/data-driven, and AI-driven development. While the second step is similar to the last step of the Stairway to Heaven model, the last step enhance in this regard by the use of ``artificial intelligence techniques such as maching learning and deep learning''.

Tanimoto~\cite{Tanimoto1990,Tanimoto2013} proposed a hierarchy for liveness, i.e., the ability to modify a running program, in visual programming. The initial proposal~\cite{Tanimoto1990} contained four levels, and was extended with two more levels in a follow-on paper~\cite{Tanimoto2013}. The six levels are: 1) informative, 2) informative and significant, 3) informative, significant, and responsive, 4) informative, significant, responsive, and live, 5) tactically predictive, and 6) strategically predictive. In the first level, the visual representation of the program is only used by the developer for comprehending and documenting the program~\cite{Tanimoto1990}. In the second level, the representation is the actual implementation of the program and will be used to perform the computation. In the third level, any change to representation triggers the execution or re-execution of the associated parts of the system after a period of time. In the level 4, the system would not wait and would be updated as soon as changes were made. In the level 5, the system is able to predict the next developer action, possibly suggesting multiple alternatives, planning ``ahead slightly to discover possible nearby program version'' useful for the developer~\cite{Tanimoto2013}. Finally, in the level 6, the system would be able to make predictions regarding a larger unit of software, sythesizing information from the software behavior and ``a large knowledge base''~\cite{Tanimoto2013}.

In summary, the taxonomies used in SE were either: 1) imported from other fields, such as Human-Computer Interaction~\cite{Feldt2018}, without proper consideration of the idiosyncrasies of SE, or 2) they were limited in the scope as the one of Savary‐Leblanc et al.~\cite{SavaryLeblanc2023} that considers only a small number of types. For example, assistants with different levels of automation could be classified as active recommender systems. Therefore, there is a clear need of an improved taxonomy, focused on SE tasks, taking into consideration the last developments in the field of LLMs and generative AI in general.

Taxonomies for the automation of tasks have also been proposed in other  fields. A notorious example is the taxonomy for driving automation systems~\cite{ISO22736} proposed by SAE International and recognized as a standard by ISO. The document describes six levels of driving automation: the first level (0) describes no automation; Level 1 regards when there is driver assistance. Level 2 stands for partial driving automation; Level 3, conditional driving automation, Level 4, high driving automation, and, finally, Level 5, which regards full driving automation. The classification is based on how much the driver participates in the process. Even though it is simple, consisting of only one dimension, the taxonomy has been employed by the industry and governments~\cite{Zanchin2017}.

\section{Research method}
\label{sec:research_method}

Taxonomies can be created based on several research methods, either as secondary studies or primary studies, employing diverse data collection methods, such as interviews, observation, questionnaires, or archival analysis~\cite{Usman2017,Ralph2019}. Secondary studies are considered one of the best options to generate a taxonomy, given its speed, connection with existing theories, and greater rigor~\cite{Ralph2019}. A natural choice would be to conduct a systematic review or mapping study in SE literature to identify tools that automate the processes. However, the SE literature breadth and the interval time that should be considered, given the goal of considering all attempts for automating the different tasks, would lead to a massive number of papers to be evaluated. Ralph~\cite{Ralph2019} suggest that taxonomies can be created ``by synthesizing, combining, refining or adapting existing theories from SE or reference disciplines.'' Given the success of the taxonomy for autonomous driving, suggested by its adoption by industry and governments~\cite{Zanchin2017}, we decided to take it as a reference to build a similar taxonomy to SE automation. 

To perform this adaptation, we performed a literature review focusing specially on secondary studies that classified, to some extent, automation tools for SE, as described in Section~\ref{sec:related_work}. We also reviewed primary studies, especially recent ones that focused on novel solutions, such as ChatGPT, which, given the short time interval since the launch, would probably not be reported in scientific published secondary studies. Based on this process, we propose a taxonomy for the automation degree of SE tasks described below.

To evaluate the taxonomy, we identified some tools regarding two SE activities, namely coding and testing, and classified them according to the taxonomy's classes. We aim to demonstrate that it fulfills the criteria suggested by Ralph~\cite{Ralph2019}, i.e., 1) its class structure should reflect ``similarities and dissimilarities between instances'', 2) its classes facilitate the inference of properties of an instance based on how it has been classified, and 3) it is effective regarding the proposed purpose. Considering different approaches for taxonomy evaluation~\cite{Usman2017}, this evaluation study can be classified as an utility demonstration, in which the taxonomy's utility is demonstrated by actually classifying subject matter examples~\cite{Smite2014,Wheaton1968}. 

\section{DAnTE: A Taxonomy for SE automation}
\label{sec:taxonomy}

\begin{figure}[ht]
\includegraphics[width=\textwidth]{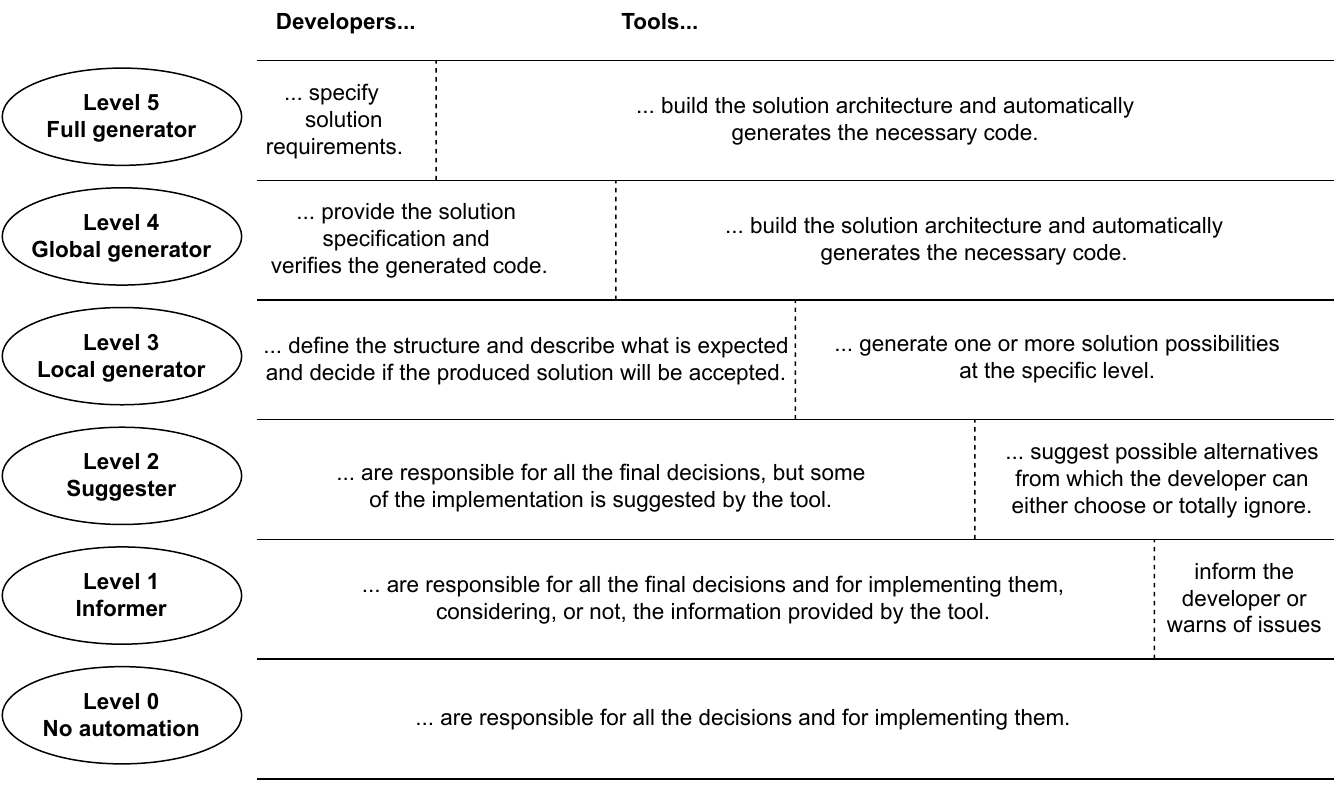}
\caption{DAnTE and its six levels of automation degree.} \label{fig:taxonomy}
\end{figure}

DAnTe, a Degree of Automation Taxonomy for SE, consists of six levels, from a level in which no automation is employed (Level 0) to a level with complete automation (Level 5). As the level increases, on the one hand, tools are able to perform more tasks with a greater level of autonomy, and, on the other hand, the responsibilities of developers decrease. Fig.~\ref{fig:taxonomy} summarizes all the levels. Dante Alighieri (c. 1265-1321) was an Italian poet who wrote the \textit{Divine Comedy}, the greatest literary work in the Italian language. In the \textit{Comedy}, Dante describes the Hell, the Purgatory, and the Heaven in different levels each.  

\subsection{Level 0 - No automation}

This level considers the scenario with a total lack of automation or support tools. The typical example is a developer or development team employing simple text editors to create code. This scenario represents how software was developed in the early days of computing, especially before the advent of integrated development environments (IDEs)~\cite{Teitelman1981}. Except for educational settings, we do not expect to find professional software being developed this way nowadays.

\textbf{Developer involvement.} The developer is completely responsible for performing the tasks and making all the decisions required for achieving the goals.

\textbf{Tool involvement.} None.

\subsection{Level 1 - Informer}

This level represents the situations in which automatic tools are able to provide developers useful information, however they are not able to suggest improvements. Some examples are tools included in simple IDEs that manage to check the code syntax and warn the user of problems, but are not able to propose fixes. More recent tools in this level are tools that automatically identify software vulnerabilities and warn developers of such issues. 

\textbf{Developer involvement.} The developer is responsible for performing the task and making all the decisions needed considering, or not, the information provided by the tool.

\textbf{Tool involvement.} The tool informs the developer or warns of issues, but is not able to produce suggestions for fixing them.

\subsection{Level 2 - Suggester}

At this level, tools are able not only to identify issues but also to propose solutions for the developers. In other words, the tool is able to automatically suggest modifications to the code originally created by the developer. Besides that, the decision to incorporate or not the suggestions is made by the developer. There are several examples in modern IDEs, such as auto-completion tools or refactoring suggesters. Another example is DependaBot\footnote{https://github.com/dependabot}, a tool for GitHub that automatically identifies dependencies for the project in a repository that could be updated and suggests the most recent versions by automatically updating dependencies files.

\textbf{Developer involvement.} The developer is responsible for performing the tasks, but the tool is able to, automatically or not, suggest modifications.

\textbf{Tool involvement.} The tool suggests one or more possible alternatives from which the developer can either choose one or, even, ignore all of them.

\subsection{Level 3 - Local generator}

At this level, tools are able to automatically perform a task in a constrained situation, with limited scope, generally based on a brief description provided by the developer. At a module, component, or solution level, the developer, or development team, is still responsible for the conception. An example is the GitHub Copilot that, at the moment of writing, is able to compose methods, functions, or even classes, based on comments or function signatures, however, the overall structure of the code is defined by the developer. 

\textbf{Developer involvement.} Developers define the structure at a certain level and describe either in natural language or a domain-specific language what is expected. They decide if the proposed solution will be accepted or, in case many possibilities are given by the tool, which one will be used. In some situations, the developer might also fix issues or adapt the solution to other parts of the system. For example, in the case of code generation, the developer could perform variable renaming to adapt the names to other parts of the system.

\textbf{Tool involvement.} Given a description for a task with limited scope, the tool is able to generate one or more possibilities of solutions.

\subsection{Level 4 - Global generator}

Tools at this level might be able to produce complete solutions given a description in a natural or domain-specific language. However, a developer still needs to verify and, if needed, modify the solution. The tools could recognize that specific parts of the system are beyond their capabilities and recommend that developers should implement these parts. Currently, some studies have evaluated the use of LLMs, such as ChatGPT, to achieve this level, with still modest results. Other tools at this level are limited to specific scenarios, such as no-code tools.

\textbf{Developer involvement.} The developer only provides the solution specification and verifies the generated solution.

\textbf{Tool involvement.} The tool implements a solution, including any necessary architecture and code.

\subsection{Level 5 - Full generator}

In this final level, tools would be able to develop a complete solution based on a set of specifications without further human intervention. Currently, some no-code tools might be able to perform similar tasks in really restricted contexts.

\textbf{Developer involvement.} The developer only needs to specify the solution requirements and input them either as natural or a domain-specific language to the tool. 

\textbf{Tool involvement.} The tool builds the solution architecture and automatically generates the necessary code.
    
\section{Evaluation}
\label{sec:evaluation}

To evaluate DAnTE, we performed an utility demonstration~\cite{Usman2017}, i.e., we applied it to classify tools and approaches on two SE activities: coding and testing.

\subsection{Coding}
\label{sec:evaluation_coding_tools}

\subsubsection{Level 0.} This scenario consists of developers using simple text editors, without any specific features for handling code, such as syntax checker, to write code. In the early days of software development, around the 1960s, until the beginning of the 1980s, programs were small and generally done as an individual effort~\cite{Shaw1990}. The need for larger systems and development teams led to the emergence of development environments and integrated tools~\cite{Shaw1990}. Since then, it is hard to imagine software being produced without any supporting tool, except for small educational tasks. 

\subsubsection{Level 1.} In this level, we include tools or simple code editors that, for instance, are able to identify syntax errors or the use of non-existent variables or functions. A contemporary example is the use of linters, ``lightweight static analysis tools that employ relatively simple analysis to flag non-complex programming errors, best practices, and stylistic coding standards.''~\cite{FerreiraCampos2019} These tools are especially useful for dynamically typed languages to warn about errors that, in these languages, would be only observed in runtime. ESLint\footnote{https://eslint.org/} for JavaScript and PyLint for  Python\footnote{https://www.pylint.org/} are some examples. 

\subsubsection{Level 2.} In this level, tools leverage similar scenarios in a limited context to make suggestions of modifications on the code for the developer. These processes are implemented using heuristics relying on similarity. A clear example is auto-completion tools in IDEs. Once the developer starts typing, the tool looks for similar terms from the language or elements in the project that could complete what the developer wanted to do. Another example could be automatic refactoring that, for example, has been for long supported by the Eclipse IDE for Java~\cite{Xing2006}. In this case, developers select pieces of code and which type of refactoring they wish to perform, the tool, then, suggests the refactoring based on pre-defined rules, and the developer can accept or not the suggested modifications.

\subsubsection{Level 3.} Tools labeled in this level are able to generate code for a given limited scenario, generally restricted to a few lines of code and few degrees of freedom. An example is GitHub Copilot's that creates the code of a function based on a comment or the signature provided by the developer. The developer can also select one of many solutions suggested by the tool. GitHub Copilot's suggestions have low complexity code that could be further simplified or that relies on undefined helper methods~\cite{Nguyen2022}. It has been also observed that it struggles to combine methods to form a solution~\cite{MoradiDakhel2023}.

\subsubsection{Level 4.} Starting from this level on, at the time of writing, no tools fulfill the needs in a reliable way. A tool in this level should be able to, given a description, develop the overall solution, including the design and implementation. In a object-oriented scenario, it would propose the classes, their interactions, and also the implementation of the methods. However, the tool would not guarantee the correctness and developers should still check the proposed design and implementation. Several researchers and practitioners have experimented with ChatGPT to investigate whether the tool is able to support the development of a software system architecture, e.g., \cite{Ahmad2023}, however the results at the time of writing did not support the capacity of the tool to fulfill this goal.

\subsubsection{Level 5.} At this final stage of the evolution of automatic generation, tools should be able to completely and reliably develop a software solution based on a natural language description of the required features. As in the driving automatic scenario in which last level it is only needed to define the destination, code generation tools in Level 5 simply require the goals to generate the final solution. The advancements of LLMs might lead to this stage in medium or long-term. Another set of tools hinting how these tools could be are low-code/no-code platforms~\cite{Sahay2020,Rokis2022}. By leveraging model-driven engineering, automatic code generation, cloud infrastructures, and graphical abstractions, these tools allow the development of fully functional applications by end-users with no particular programming background, the so-called citizen developers~\cite{Sahay2020}. However, these platforms require that the users model the domain, define the user interface, and specify business logic~\cite{Sahay2020}. Level 5 tools would leverage natural language processing to perform these steps automatically based on, for example, a chat with users.

\subsection{Automated Testing}
\label{sec:testing_tools}

\subsubsection{Level 0.} This scenario consists of developers elaborating the code of automated tests manually, using only the requirements and information about the API that should be used for testing. Only a testing framework, such as JUnit \cite{cheon2002simple}, is used to provide some structure to the tests. The setup for the tests and the verifications are configured through code. The developer has no aid about the test structure and its coverage. 

\subsubsection{Level 1.} At this level, the environment can provide some additional information about the tests that can guide the developer to create the tests and the test code. An example of a feature present in some tools that can help for a better test code is automated test smell detection \cite{palomba2018automatic,peruma2020TsDetect}. This information can guide the developer to create a cleaner code and a more efficient test suite. Another useful information that can be provided is code coverage~\cite{shahid2011evaluation}, which is usually shown visually in the IDE through the lines of code being covered by the tests or not. Based on that, the developer might add new tests or modify the existing ones to achieve the desired coverage. 

\subsubsection{Level 2.} In this level, we already have suggestions to create or improve the tests. Similar to coding, there are suggested refactorings specific to test code that can be applied automatically by the tools \cite{xuan2016b,marinke2019towards}. Another kind of tool that fits into this level is the creation of test templates. Most IDEs provide the functionality to create a test class template based on the structure of the class to be tested, suggesting the test methods that should be created. The test template generation might also consider other factors, such as the test names~\cite{zhang2015automatically}. 

\subsubsection{Level 3.} At this level, we have the generation of test code for limited and specific contexts, like unit tests of isolated classes and methods, e.g., \cite{mcminn2004search}. These tools usually consider the internal structures of the test target, generating test suites that try to maximize the coverage~\cite{fraser2011evosuite,braione2018sushi}. Some preliminary research also explored the usage of LLMs, based on ChatGPT, for unit test generation~\cite{yuan2023no} with promising results. Even if some results indicate that the coverage of generated test suites is similar to manually created tests~\cite{kracht2014empirically}, other studies show that several faults were not detected by these automatically generated tests~ \cite{shamshiri2015automatically}. For instance, these tools have difficulties generating tests with good coverage when the inputs need a specific format or type to enter a given code branch. 

\subsubsection{Level 4.} At this level of the taxonomy, the tests would be created not for a class or a method but considering the whole application. In that scope, we can find tools that generate functional tests for specific kinds of applications, like web apps~\cite{dallmeier2014webmate} and REST APIs~\cite{arcuri2017restful}. For more specific cases, it is also possible to find test generation for non-functional aspects, like a work that proposed the generation of tests to detect leaks in Android applications~\cite{zhang2016automated}. Even if we have some test-generation tools that could be classified at this level for targeting the application as a whole, they are still specific to platforms and focus the tests on a single aspect. 

\subsubsection{Level 5.} For the full generation of automated tests, the tool is expected to evaluate the application structure and generate tests at different levels. Additionally, its scope should not be limited to functional tests, also considering other non-functional aspects, such as security and performance. Currently, there is no tool that reaches this degree of automation for test code generation.   

\section{Discussion}
\label{sec:discussion}

The taxonomy proposed in this chapter, DAnTE, aimed to increase cognitive efficiency and facilitate inferences, as is the general case for taxonomies~\cite{Ralph2019}, for the specific case of automation of SE tasks. The validation performed through an utility demonstration, i.e., by classifying coding and testing tools, demonstrated that different tools, even for distinct tasks, share similarities and that the taxonomy could be employed to group them in categories, facilitating comparison. 

Comparing with the existing taxonomy for the application of AI for SE tasks, i.e., AI-SEAL~\cite{Feldt2018}, and the classification of assistants for SE~\cite{SavaryLeblanc2023}, DAnTE has some advantages. It has less steps when compared to AI-SEAL and they are similar to those of a taxonomy of another field, i.e., driving automation. This aspect facilitates the comprehension by practitioners also by giving an analogy. In particular, the taxonomy development was based on the parsimony principle, that favors the simplest explanation possible~\cite{Gori2024}, as the reduced number of levels show. Besides that, the proposed steps were redesigned with concepts and terminology related to SE, and not simply adapted from another field, as human-computer interaction. The number of steps allows, however, to differentiate tools that in the classification proposed for software assistants would be grouped together. Most, if not all, tools classified in the levels above 2 would be classified as active recommender systems even though, as our examples show, they have very different capabilities and degree of automation.  

When classifying some practices in the levels 4 and 5, we could observe that some tools are probably going to be able to perform diverse tasks, such as, coding and testing. Therefore, a possible further development of the taxonomy could consider the capacity of a tool of fulfilling multiple tasks rather than, for example, the point of application of the AI-SEAL taxonomy.

Ralph~\cite{Ralph2019} identifies five possible errors when creating taxonomies: inclusion, exclusion, combination, differentiation, and excess. On one hand, an inclusion error happens when an instance is included in a class even though it is not similar to other instances of the class. On the other hand, an exclusion error occurs when an instance is not grouped with other instances similar to it. In the utility demonstration we performed, we presented the differences among the tools that made us grouped them together or apart. Combination errors happen when not similar instances are grouped together. In this regard, DAnTE is an improvement regarding the existent classification of software assistants, in the sense that it differentiates tools with different degrees of automation. Differentiation errors represent the cases when similar instances are separated in different classes. Based on our discussion above, we argue that the proposed taxonomy avoids the possibility of this error that could happen when using AI-SEAL and its ten levels of automation by reducing the number of steps and providing an analogy. Finally, an excess regards cases when a class has no instances. Although the Level 5 might fit this definition, it is useful to guide the development of new tools that, based on the rapid development of generative AI and LLMs, we expect to have them available in the near future.

\section{Conclusions}
\label{sec:conclusions}

This chapter presented DAnTE, a taxonomy for the automation degree of SE tasks, by leveraging a taxonomy for driving automation and reviewing similar works for SE. We performed an initial evaluation based on utility demonstration, i.e., by classifying coding and testing tools, reaching promising results. The taxonomy could help the comparison of diverse tools. It could also help practitioners to understand what they could expect from different tools and facilitate the decision process of tool selection. The taxonomy also hints the next steps of automation of SE tasks, guiding practitioners and researchers alike in the development of novel tools.

Given the limited space, we were not able to present the use of the taxonomy for other SE tasks. However, our preliminary results indicate that it is useful to classify diverse tools. A comprehensive analysis might indicate that the emergence of Level 4 or 5 tools could lead to the merge of different tasks as our evaluation showed, i.e., coding tools that already test the code produced. The taxonomy could be also further developed, probably including other dimensions, such as the type of AI applied. For example, the capability of a tool of performing more than a task, as described above, might indicate a valuable dimension for the classification of these tools. As the generative AI technology advances and new tools for SE tasks are proposed, future work could investigate the addition of new dimensions for the taxonomy.

 \bibliographystyle{splncs04}
 \bibliography{refs}

\end{document}